\documentclass[10pt]{article}

\usepackage[OE]{express}
\usepackage{graphicx}
\usepackage{braket}
\usepackage{amsmath}
\usepackage{amssymb}
\usepackage{gensymb}
\usepackage{epstopdf}
\usepackage[normal]{subfigure}
\usepackage{hyperref}
\begin{document}

\title{Repeated Measurements with Minimally Destructive Partial-Transfer Absorption Imaging}

\author{Erin~Marshall~Seroka,\authormark{1,*} Ana~Vald{\'e}s Curiel,\authormark{1}, Dimitrios~Trypogeorgos,\authormark{1} Nathan~Lundblad,\authormark{1,2} and Ian~B.~Spielman\authormark{1}}

\address{\authormark{1}Joint Quantum Institute, University of Maryland and 
National Institute of Standards and Technology, College
Park, Maryland, 20742, USA\\
\authormark{2} Department of Physics \& Astronomy, Bates College, Lewiston, Maine, 04240, USA}

\email{\authormark{1}erinmars@physics.umd.edu}

\begin{abstract}
We demonstrate partial-transfer absorption imaging as a technique for repeatedly imaging an ultracold atomic ensemble with minimal perturbation. We prepare an atomic cloud in a state that is dark to the imaging light. We then use a microwave pulse to coherently transfer a small fraction of the ensemble to a bright state, which we image using \textit{in situ} absorption imaging. The amplitude or duration of the microwave pulse controls the fractional transfer from the dark to the bright state. For small transfer fractions, we can image the atomic cloud up to 50 times before it is depleted. As a sample application, we repeatedly image an atomic cloud oscillating in a dipole trap to measure the trap frequency. 
\end{abstract}


Ultracold atoms are a near-ideal platform for studying quantum many-body physics. The combination of well understood control and measurement techniques allow for quantitative comparison between experiment and \textit{ab initio} theory.  In most experiments, the atomic ensemble is directly imaged in a destructive manner, requiring the preparation of a new ensemble for each measurement. The most common technique is resonant absorption imaging after a time-of-flight (TOF) expansion, in which the atoms are released from the confining potential and imaged using a resonant probe beam. The primary disadvantage of absorption imaging is that the photon recoil in the atom-light scattering process imparts significant kinetic energy to the atoms. As such, the absorption imaging process destroys the atomic cloud, and only one image can be captured every time an atomic ensemble is prepared. For experiments in which certain information is needed, such as the center of mass position of the cloud, data acquisition can be considerably accelerated by repeatedly imaging the same cloud rather than creating a new ensemble for every data point. The desirability of more efficient data collection motivates the development of imaging techniques that minimally perturb the atomic ensemble.

%% FIGURE 1
\begin{figure}[ht]
	\begin{center}	\includegraphics[width=0.8\linewidth]{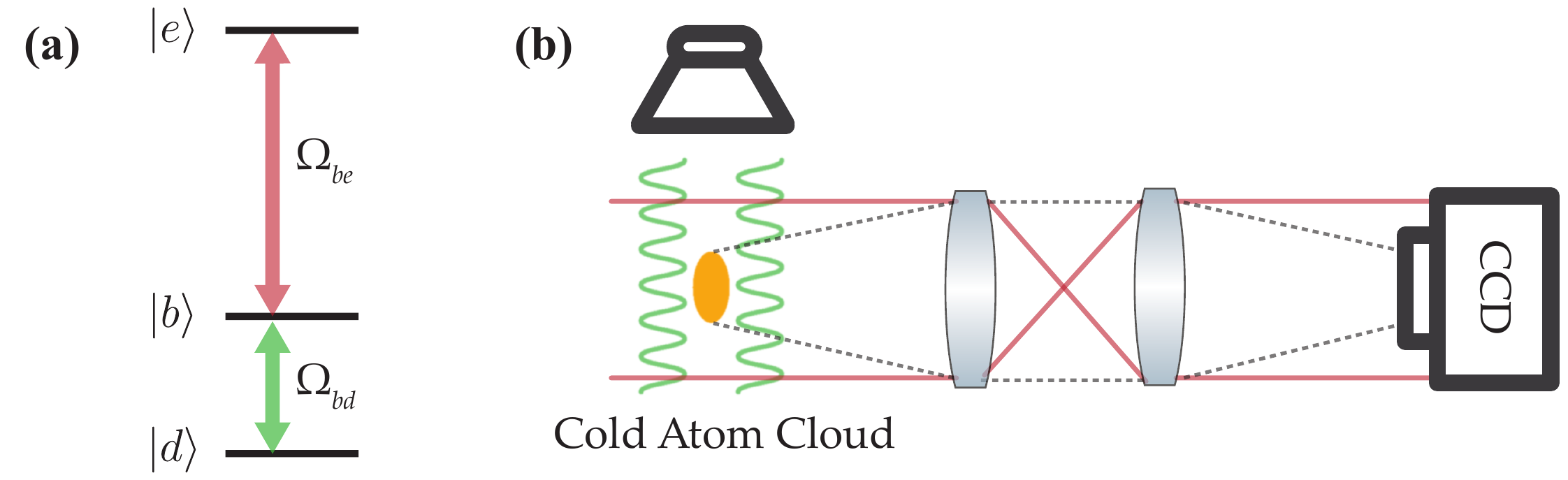}
		\caption{  \textbf{(a)} Level diagram for the partial-transfer absorption imaging technique. The atoms are prepared in the dark state $\ket{d}$, and a fraction of the atoms are transferred from $\ket{d}$ to $\ket{b}$ by a coherent pulse (green) of frequency $\Omega_{bd}$. \textbf{(b)} A simple schematic of the partial-transfer absorption imaging system. The pink lines depict the probe beam, and the dotted region indicates a small solid angle of the light of frequency $\Omega_{be}$, emitted isotropically from the cloud as the atoms decay from $\ket{e}$ to $\ket{b}$. The light is collected by the imaging lenses and re-focused onto a charge-coupled device (CCD).}
		\label{fig0}
	\end{center}
\end{figure}

Most imaging techniques capable of repeatedly sampling the same ensemble are dispersive; the portion of the probe beam that interacts with the atoms receives a phase shift due to the high index of refraction of the atomic cloud. For off-resonant laser light, the polarizability describing the light-matter interaction can be divided into scalar, vector and tensor contributions \cite{Jasperse2015}, of which the scalar and vector are commonly applied to dispersive imaging.

The scalar polarizability imparts a phase shift proportional to the atomic density, but independent of the internal atomic state.\footnote{Because the phase shift is also dependent on the detuning of a transition from resonance, even the scalar polarizability can be made state-dependent.} In the original implementation, called dark-field phase-contrast imaging \cite{Andrews1996}, a small opaque beam block was placed in the imaging path at the focal plane of the laser to block the unperturbed laser light. The only light that passed to the detector was the light that acquired a phase shift, and the intensity of the signal was proportional to the square of the atom-induced phase shift. In later implementations of phase-constrast imaging, the opaque object was replaced by a transparent phase dot, which imparts a phase shift of $\pi/2$ to the probe beam \cite{Andrews1997}, greatly increasing the signal-to-noise ratio of the images. In a similar technique, phase-contrast polarization imaging \cite{Bradley1997}, a linear polarizer causes the probe beam and the scattered light from the atoms to constructively interfere, creating an image of the atomic cloud on the camera.

Alternatively, the light might acquire a vector phase shift that is proportional to the atoms' internal angular momentum, and dependent on the polarization state of the probe light. For a linearly polarized probe beam, the line of polarization rotates. An example of a technique that makes use of the vector phase shift is Faraday imaging \cite{Gajdacz2013}, in which a polarizing beam-splitter cube between the atomic cloud and the camera separates out the horizontal and vertical polarization components. The polarization of the probe beam is set such that the transmission through the cube is minimized in the absence of atoms. When
atoms are present, the polarization rotation imparted to the beam allows some
light to pass through to the camera. Similarly, in dual-port Faraday rotation imaging \cite{Kaminski2012}, the two outputs of the beam-splitter cube are projected side-by-side onto a camera. A comparison of the two images allow for calculation of the cloud column density and spatial distribution. Most dispersive techniques rely on either the scalar or vector polarizability, although some, such as partial phase-contrast imaging \cite{Wigley2016}, use a combination of the two.

Non-dispersive methods for minimally-destructive imaging have also been
implemented. For example, fluorescence imaging with photodetectors operating at the shot-noise limit \cite{Lye2003} detects the light scattered from the atomic cloud, but the acquired signal is typically weak due to the near-isotropy of the scattered light and the small solid angle over which that light can be collected. Diffraction-contrast imaging \cite{Turner2005} uses the diffraction of the probe beam due to the atoms to image the cloud.

Partial-transfer absorption imaging (PTAI), on the other hand, is a resonant absorption imaging technique that images only a small fraction of the cloud at a time, allowing for the same cloud to be repeatedly sampled. A fraction of the atoms in the atomic ensemble are transferred by a microwave pulse to an internal state that is on a cycling transition. The transferred atoms are imaged while the rest of the atoms remain virtually invisible to the imaging probe beam. The recoil momentum from the probe beam ejects the imaged fraction from the trap, and another fraction of the cloud may be imaged in the same way as the previous fraction. An early implementation of multi-shot absorption imaging transferred a fraction of atoms into a magnetically untrapped state, where they were then imaged in time-of-flight \cite{Freilich2010}. The first demonstration of PTAI for \textit{in situ} applications showed that partial-transfer can be used to optimally image clouds of high optical depth by sampling a fraction of the cloud \cite{Ramanathan2012}. Imaging a fraction of the cloud is a repeatable step, and the fraction imaged may be tuned, but the range of possible perturbations to the cloud has not yet been characterized.

Here, we demonstrate that the fraction of the cloud that is imaged may be easily tuned using either the duration or the amplitude of the microwave pulse, showing that PTAI is a tuneable minimally-destructive imaging technique well-suited for repeatedly imaging the same ensemble. PTAI is advantageous over other techniques in that the degree of perturbation to the cloud can be easily varied without the need to manufacture a phase dot, construct parallel beam paths, align a beam-splitter cube, or find an optimal probe beam detuning.

This manuscript is organized as follows. In Section \ref{tech}, we discuss the theory underlying partial-transfer absorption imaging and define a way of quantifying the reproducibility of the measurement. In Section \ref{exper}, we give more detail about our implementation of PTAI for imaging $^{87}$Rb Bose-Einstein condensates (BECs). In Section \ref{demtune}, we show how the transferred fraction can be tuned to select the number of images captured and their relative signal. Finally, in Section \ref{dipolemode}, we demonstrate an example application of the multi-shot PTAI technique: measuring a collective oscillation of a trapped atomic cloud.

\section{Technique} \label{tech}

Conventional absorption imaging derives the column density of an atomic cloud from three images. In the first image, the atoms are exposed to an imaging probe beam that is resonant with an atomic transition. Light is absorbed and rescattered by the atomic ensemble, locally depleting the probe laser and leaving a shadow in the image of the beam as it arrives at the camera with intensity ${I}_{\mathrm{atoms}}$. The second image is taken under the same conditions as the first but with the atoms absent, to give an intensity ${I}_{\mathrm{probe}}$ due to the probe beam. In the third image, the probe is turned off and ${I}_{\mathrm{bgnd}}$ comprises the dark count noise of the camera, along with light from all other sources. The background intensity is subtracted from the other two intensity distributions to give ${I}_{f}={I}_{\mathrm{atoms}}-{I}_{\mathrm{bgnd}}$ and ${I}_{i}={I}_{\mathrm{probe}}-{I}_{\mathrm{bgnd}}$.

%% FIGURE 2
\begin{figure}[ht]
	\begin{center}
\includegraphics[width=0.8\linewidth]{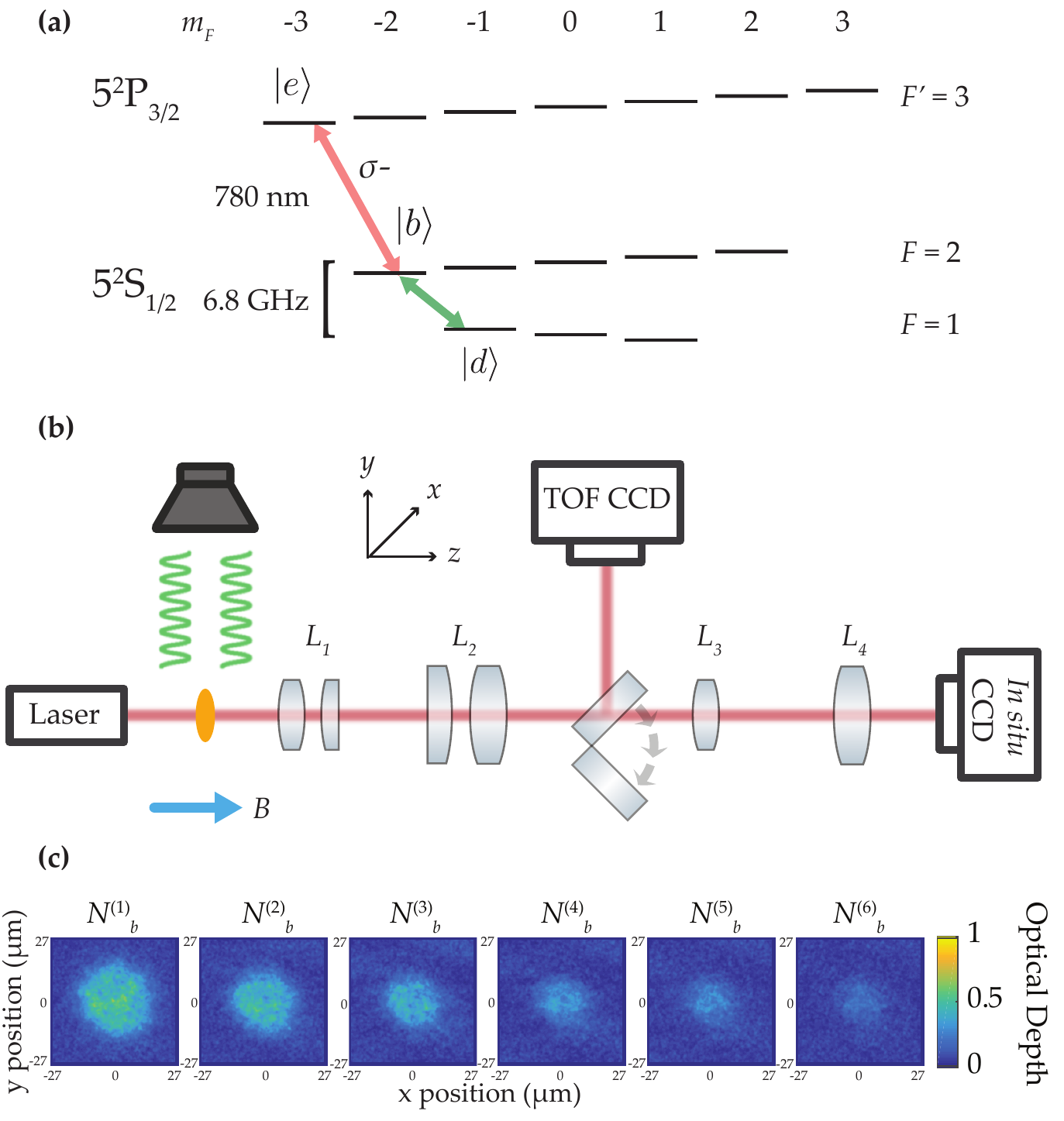}
		\caption{  Partial-transfer absorption imaging. \textbf{(a)} We used the hyperfine structure of the $5^{2}S_{1/2}$ electronic ground state of $^{87}$Rb to resolve the $\ket{d}$ and $\ket{b}$ states, and we imaged on the D$_{2}$ transition. A 6.8 GHz coherent microwave pulse (green) coupled the dark state, $\ket{1,-1}$, to the bright state, $\ket{2,-2}$. A $\lambda =$ 780 nm laser (pink) coupled the bright state to the excited state, $\ket{3,-3}$. \textbf{(b)} The imaging system, including the TOF and \textit{in situ} imaging paths, is shown. $L_{1}$ is a compound lens pair consisting of a 60 mm achromatic doublet lens and a 300 mm plano-convex singlet lens. $L_{2}$ is a second compound lens pair consisting of a 500 mm plano-convex singlet and a 250 mm achromatic lens. $L_{3}$ is a 150 mm achromatic doublet. $L_{4}$ is a 750 mm achromatic doublet. \textbf{(c)} Six PTAI images taken of the same BEC using the \textit{in situ} CCD camera as shown. For each image, a 12 \micro s pulse at a Rabi frequency of approximately 14.3 kHz transferred a fraction of the cloud from the dark state to the bright state. The atoms were imaged on the cycling transition.}
		\label{fig1}
	\end{center}
\end{figure}

The change in beam intensity as the photons are absorbed, which carries information about the cloud column density, is described by Beer's Law. The imaging beam is attenuated along the optical axis ${\bf e}_z$ as it passes through an atomic cloud with number density $n(x,y,z)$. Including saturation effects present when the imaging beam's intensity ${I}$ is not negligible compared to $I_{\mathrm{sat}}$, where $I_{\mathrm{sat}}$ is the saturation intensity for the relevant atomic transition, Beer's law is

\begin{equation} \label{Beer}
\frac{d\widetilde{I}}{dz} = -n \frac{\sigma_{0}}{\alpha} \frac{\widetilde{I}(z)}{1+\widetilde{I}},
\end{equation}

\noindent where $\sigma_{0}$ is the resonant cross section \cite{Genkina2016,Reinaudi2007}. Here $\widetilde{I} = I/I_{\mathrm{sat}}$, such that all intensities are normalized to $I_{\mathrm{sat}}$. The parameter $\alpha$, which must be determined experimentally, accounts for experiments imperfections, such as imperfect polarization of the probe beam, and corrections in approximating a multi-level atom as a two-level system.

Beer's law (Eq. \ref{Beer}) can be integrated to give the optical depth (OD), or the integrated density along the imaging axis, as

\begin{equation} \label{nsig0}
\sigma_{0} n_{2D} = - \alpha \: \mathrm{ln} \left( \frac{\widetilde{I}_{f}}{\widetilde{I}_{i}} \right) + \widetilde{I}_{i} - \widetilde{I}_{f},
\end{equation}

\noindent in terms of $\widetilde{I}_{i}$ and $\widetilde{I}_{f}$, the intensities just prior to and just following the atomic ensemble, respectively.

In PTAI, the trapped atoms are prepared in an electronic ground state $\ket{d}$ that is dark to the imaging light, as shown in Fig. \ref{fig0}. Before an image is taken, a fraction $\epsilon$ of the atoms is transferred from $\ket{d}$ to a bright state $\ket{b}$ using a coherent coupling field with Rabi frequency $\Omega_{\rm bd}/2\pi$ and then resonantly imaged on a $\ket{b} \leftrightarrow \ket{e}$ cycling transition, where $\ket{e}$ is an electronically excited state. The large radiation pressure from the imaging light serves to expel any atoms measured in $\ket{b}$ from the trap, leaving behind the remainder in $\ket{d}$ ready to be imaged once more. For small $\epsilon$, the same atomic cloud may be repeatedly sampled until it is slowly depleted by the succession of weak measurements.

Since most experiments capture image data on a CCD or a complementary metal-oxide-semiconductor (CMOS) sensor, we consider the number of atoms imaged onto a pixel of area $A$. The total number of atoms in the frame is the sum of the atom number in all pixels $N = \sum_{i} n_{2D} A$. The number of atoms expected in a PTAI image is the number of atoms that were transferred into $\ket{b}$. The atoms in the bright state represent a fraction $\epsilon$ of the atoms initially prepared in $\ket{d}$, where

\begin{equation}
\epsilon = \mathrm{sin^2} \left( \frac{\Omega_{\mathrm{bd}} t}{2} \right) .
\end{equation}

\noindent Here $\Omega_{\mathrm{bd}} / 2\pi$ is the Rabi frequency, and $t$ is the duration of the coupling pulse. The number of atoms in  $\ket{d}$ before the $m^{th}$ PTAI shot is $N_{d}^{(m)}$, and the $m^{th}$ shot samples $\epsilon N_{d}^{(m)}$ atoms, so the number of atoms $N_{b}^{(m)}$ in the $m^{th}$ PTAI shot is

\begin{equation} \label{absnumber}
N_{b}^{(m)} = \epsilon N_{d}^{(1)} { \left( 1 - \epsilon \right) }^{m-1} = N_{b}^{(1)} \mathrm{cos}^{2(m-1)} \left( \frac{\Omega_\mathrm{{bd}} t}{2} \right) ,
\end{equation}

\noindent
where $N_{b}^{(1)}$ is the number of atoms in the first PTAI image. Eq. \ref{absnumber} can also be expressed as an exponential function,

\begin{equation} \label{expparam}
N_{b}^{(m)} = N_{b}^{(1)} \exp \left( -\frac{m-1}{\xi} \right), \quad \mathrm{with} \quad \xi=-1/\mathrm{ln}\left( 1-\epsilon \right)
\end{equation}

\noindent as a way of quantifying the imaging repeatability in terms of the parameter $\xi$, which we call the longevity.

The fraction $\epsilon$ of atoms transferred from the dark state $\ket{d}$ to the bright state $\ket{b}$ can be chosen by varying either the microwave pulse duration or amplitude. For $t \ll 2\pi/ \Omega_{\mathrm{bd}}$, only a small fraction of the atoms in $\ket{d}$ will be transferred to $\ket{b}$ for each image; as a result, tens of images of the same cloud may be captured, albeit with reduced signal. On the other hand, for

\begin{equation}
t \approx 2\pi \left( q+ \frac{1}{2} \right) / \Omega_{\mathrm{bd}} ,
\end{equation}

\noindent where $q \in \mathbb{N}$, $\epsilon \rightarrow \mathrm{1}$ and only a couple images with a discernably-sized cloud may be captured. The $\epsilon \rightarrow \mathrm{1}$ limit approaches standard \textit{in situ} absorption imaging. Therefore, with multi-shot partial-transfer absorption imaging, any atomic species with a $\ket{d}$, $\ket{b}$, and $\ket{e}$ manifold like the one pictured in Fig. \ref{fig0}a may be imaged repeatedly with a tuneable degree of repeatability and signal.

\section{Experiment} \label{exper}

We demonstrated the multi-shot PTAI technique using the D$_{2}$ transition hyperfine structure of $^{87}$Rb. We began each experiment with a BEC in the $5^2S_{1/2}$ electronic ground state in the $\ket{F=1,m_{F}=-1}$ sublevel, as shown in Fig. \ref{fig1}a. The $\ket{1,-1}$ state served as our dark state $\ket{d}$. The BEC was confined in a crossed optical dipole trap formed by two $1064$ nm beams propagating along $\mathbf{e}_{x}$ and $\mathbf{e}_{y}$. We broke the degeneracy of the three $m_F$ magnetic sub-levels in the $F=1$ manifold with an approximately 2 mT bias field along $\mathbf{e}_{z}$ that produced a Zeeman splitting of $g_{F} \mu_{\mathrm{B}} B/h \approx$ 14 MHz between $\ket{1,-1}$ and $\ket{1,0}$, where $g_{F}$ is the hyperfine Land{\'e} $g$-factor, $\mu_{\mathrm{B}}$ is the Bohr magneton, $B$ is the magnetic field strength, and $h$ is Planck's constant.

%% FIGURE 3
\begin{figure}[htp]
	\centering
\includegraphics[width=0.9\linewidth]{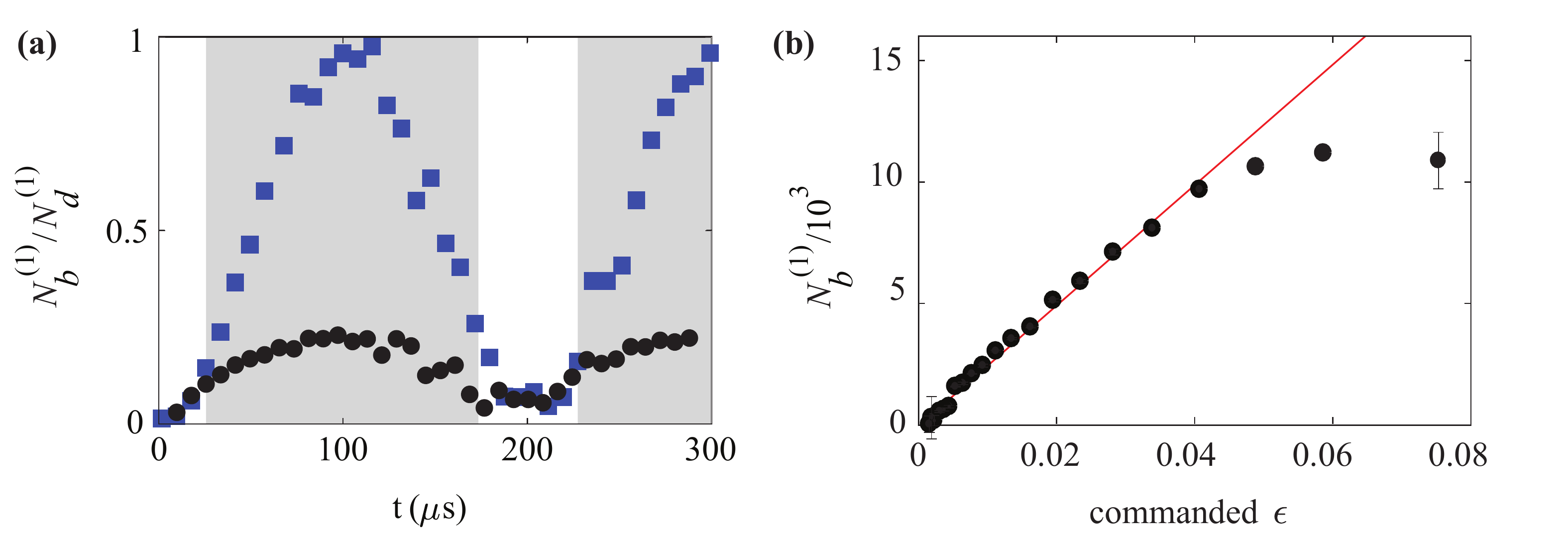}
	\caption
	{   Coherent transfer measured using TOF and \textit{in situ} absorption imaging. \textbf{(a)} The fraction of atoms transferred into $\ket{F=2}$ after variable microwave pulse duration shows Rabi oscillation with a frequency $\Omega_{\mathrm{bd}}/2\pi = 5.03(33)$ kHz. We measured the transfer fraction using TOF absorption imaging (blue squares) and \textit{in situ} absorption imaging (black circles). The grey shading indicates regions where the optical depth of a typical shot is $\gtrapprox$ 3. The transfer fraction in the grey regions is underrepresented due to extinction of the probe beam in high-OD atomic clouds. Atom number uncertainties are smaller than the data points. \textbf{(b)} The measured fraction transferred into the bright state increases with the microwave pulse amplitude. As $\epsilon$ is increased by strengthening the pulse, the number of atoms $N_{b}^{(1)}$ in the first PTAI shot increases linearly for $\epsilon\leq$ 0.04. Error bars represent the 95 \% confidence bounds on $N_{b}^{(1)}$.
	}
	\label{fig2}
	\end{figure}

We used microwave pulses, tuned to the 6.8 GHz ground state hyperfine splitting, with variable duration (from 2 $\micro$s to 304 $\micro$s) and variable Rabi frequency (from 15/$2\pi$ kHz to 90/$2\pi$ kHz) to place each atom in a superposition of $\ket{1,-1}$ and $\ket{2,-2}$. The microwave pulses effectively transferred a fraction $\epsilon$ of the BEC into $\ket{2,-2}$, which served as our bright state $\ket{b}$. The microwave transition is depicted by the green arrow in Fig. \ref{fig1}a.

The microwave pulses were generated by a Stanford Research Systems SG384 signal generator.\footnote{The identification of commercial products is for information only and does not imply recommendation or endorsement by the National Institute of Standards and Technology.} Coupling the microwave signal and an approximately 100 MHz signal from a Novatech 409B direct-digital synthesizer into a Marki IRW0618 mixer allowed us to vary the microwave signal across tens of MHz. The signal then passed through an analog voltage-controlled General Microwave Herley D1956 attenuator and a Microwave Amplifiers AM53 amplifier to control the signal amplitude. A MCLI CS-57 circulator-isolator prevented power from being reflected back toward the source and instead dumped any reflected signal into a Minicircuits 42 ZX47-40-S+ power detector. We maximized the power sent to the atoms by tuning a downstream Maury Microwave 1819C stub tuner, optimized to the output in the power detector. Finally, the signal was delivered to a microwave waveguide directed at the atoms.

Once a fraction of the cloud had been transferred to the bright state, we imaged those atoms at an intensity of $\widetilde{I}=0.25(2)$\footnote{All uncertainties herein reflect the uncorrelated combination of single-sigma statistical and systematic uncertainties.} on the cycling transition coupling $\ket{2,-2}$ to $\ket{F'=3,-3}$ (our excited state $\ket{e}$). We used resonant \textit{in situ} absorption imaging to measure the spatial density distribution of the atoms in $\ket{2,-2}$. Because the transition from $\ket{1,-1}$ to the allowed excited states is far-detuned from the imaging transition, the atoms in $\ket{1,-1}$ were only minimally perturbed during the imaging process.

Radiation pressure from the imaging process rapidly accelerated the measured atoms, expelling them from our comparatively shallow optical dipole trap, and leaving the unmeasured atoms all in $\ket{1,-1}$, ready to be measured once more. Atoms were repeatedly transferred to $\ket{2,-2}$ and imaged as the number of atoms in $\ket{1,-1}$ continued to decrease.

We used \textit{in situ} imaging for all PTAI experiments, but we also used standard TOF absorption imaging for preliminary calibrations, as well as to benchmark the performance of our \textit{in situ} imaging system. The TOF absorption imaging system contains two pairs of compound lenses, which provide a magnification of 3.19(3) for a 21 ms TOF along $\mathbf{e}_{z}$, and terminates at a PointGray Flea3 camera. Our \textit{in situ} imaging system complements the TOF imaging system with an additional telescope giving an \textit{in situ} magnification of 16.0(2) and terminates at a Princeton Instruments ProEM HS:512BX3 camera. A flipper mirror allows us to switch between the TOF absorption imaging camera and the \textit{in situ} PTAI camera. The two imaging laser-beam paths are shown in Fig. \ref{fig1}b, and sample PTAI images are shown in Fig. \ref{fig1}c.

Fig. \ref{fig2}a demonstrates Rabi oscillation between the dark and bright states as measured on our TOF absorption imaging system. A variable fraction $\epsilon$ of atoms, determined by the microwave pulse duration $t$, was transferred from $\ket{1,-1}$ to $\ket{2,-2}$. The two BECs were separated spatially during time-of-flight using the Stern-Gerlach technique. For each image, the number of atoms in $\ket{2,-2}$ was compared to the total number of atoms in both states. The fraction transferred as a function of time shows characteristic Rabi oscillation behavior with a Rabi frequency $\Omega_{\mathrm{bd}}/2\pi = 5.03(33)$ kHz. Similarly, the fraction of atoms in $\ket{2,-2}$ as a function of time, measured using \textit{in situ} absorption imaging, is shown in Fig. \ref{fig2}a. In the gray regions, the \textit{in situ} OD, calculated according to Eq. \ref{nsig0}, was in excess of 3, marking regions where \textit{in situ} absorption imaging fails. White regions indicate a smaller OD. It is difficult to image a high-OD cloud due to extinction of the imaging probe beam, so the atom number in the grey region inferred using Eq. \ref{Beer} greatly underestimates the actual transferred population.

Nevertheless, the atom number measured in the bright state in the first PTAI shot is linear in $\epsilon$ for small $\epsilon$, as shown in Fig. \ref{fig2}b, in which the transfer fraction is tuned by varying the microwave pulse amplitude rather than the pulse duration. We determined the total number of atoms originally prepared in $\ket{1,-1}$ by taking $N_{b}^{(m)}/\epsilon$ for each point falling along the line. We calculated that the total atom number is $2.168(5) \times 10^{5}$ atoms, which is the figure we used in the characterization of $\xi$.

\section{Tuneability} \label{demtune}

\begin{figure}[hp]
	\centering
	\includegraphics[width=0.9\linewidth]{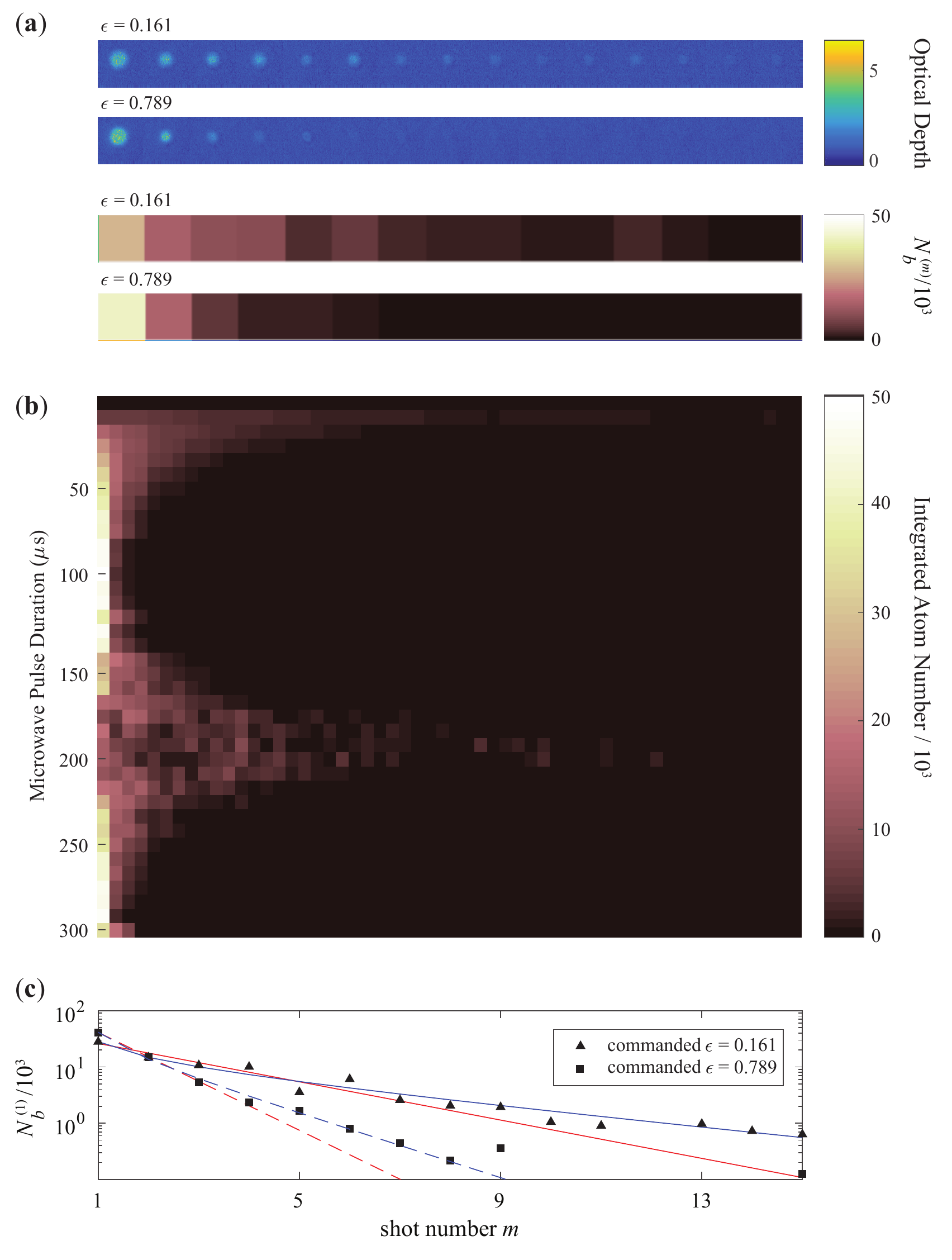}
		\caption{	The effect of transfer strength on measured atom number in a PTAI series. \textbf{(a)} Examples of weak transfer (top image) and strong transfer (second image) are shown, each displaying the first 15 images of the PTAI series. The second pair of images represent the integrated atom number in each of the 15 images for the weak transfer and strong transfer examples above. \textbf{(b)} Measured atom number varies with transfer strength, here determined by the microwave pulse duration. Each row represents the total atom number measured in each of 58 images of a single BEC, taken 4 ms apart, in chronological order from left to right, for a given pulse duration. The Rabi frequency was about 4.9 kHz. \textbf{(c)} The integrated atom number is plotted as a function of shot index for both the weak transfer series (triangles) and the strong transfer series (squares). To determine $N_{b}^{(1)}$ and $\xi$ for a given transfer fraction, the data is fit to the exponential in Eq. \ref{expparam} (red) and the rescattering model based on Eq. \ref{rescatt} (blue).}
	\label{figEXP}
\end{figure}

As in any minimally-destructive imaging system, there is a trade-off between the number of images one is able to capture of the same atomic cloud and the signal strength of each of those images. Thus, the ability to tune the system parameters to achieve a high signal and maximal longevity $\xi$ of the cloud is desirable.  As is evident from Fig. \ref{figEXP}a, choosing a high $\xi$ (low $\epsilon$) produces a long series of reduced-signal images. Choosing a low $\xi$ (high $\epsilon$) results in a series in which the first few images have enhanced signal. The system can thus be tuned to produce the desired balance of signal and number of images, as shown in Fig. \ref{figEXP}b.

%% FIGURE 4

When $\epsilon$ is tuned using either the microwave pulse duration or the microwave pulse amplitude, $\xi$ is generally observed to be smaller than expected. Atom-atom scattering is one contributing factor: as $\epsilon$ increases and more atoms in $\ket{2,-2}$ are expelled from the trap, they are likely to recoil off atoms in $\ket{1,-1}$, expelling some of the untransferred population from the dipole trap. Thus the $\ket{1,-1}$ population being sampled is smaller than predicted.

A simple model of the rescattering process suggests that the number of atoms left in $\ket{F=1}$ after the $m^{\mathrm{th}}$ image is taken is

\begin{equation} \label{rescatt}
N_{d}^{(m+1)} = N_{d}^{(m)} \frac{1-\epsilon}{ 1+\epsilon(e^{N_{d}^{(m)} \beta}-1) },
\end{equation}

\noindent rather than $N_{d}^{(m)} (1-\epsilon)$. Here $\beta$ is a dimensionless parameter that depends on the imaging pulse length, the trap volume, the scattering-cross section of the atoms, and the recoil velocity. To find $\beta$, we fit the atom number vs. shot number for many PTAI images to Eq. \ref{rescatt} with both $\epsilon$ and $\beta$ as free parameters, took a weighted average of the resulting values of $\beta$, and obtained a value of 0.0091. We used that weighted average as $\beta$ when we fit the data again to measure $\epsilon$ (and hence $\xi$) for each commanded transfer fraction.

Comparisons of the exponential model (Eq. \ref{expparam}) and the rescattering model (Eq. \ref{rescatt}) are shown in Fig. \ref{figEXP}c. Integrated atom numbers for two different PTAI image series, one with a low transfer fraction and one with a high transfer fraction, are shown. The red curves show the fit to Eq. \ref{expparam}, and the blue curves show the fit to the rescattering model in Eq. \ref{rescatt}. The reduced chi-squared values for the fits to Eq. \ref{expparam} are 0.563 ($\epsilon=0.161$) and 0.297 ($\epsilon=0.789$), and the reduced-chi squared values for the fits to Eq. \ref{rescatt} are 0.363 ($\epsilon=0.161$) and 0.292 ($\epsilon=0.789$). As one can see from the figure and the fit statistics, the rescattering model is a better fit to the data.

\begin{figure}[ht]
	\begin{center}
		$\begin{array}{ll}
		\textrm{\textbf{(a)}} & \textrm{\textbf{(b)}} \\
		\includegraphics[width=0.45\linewidth]{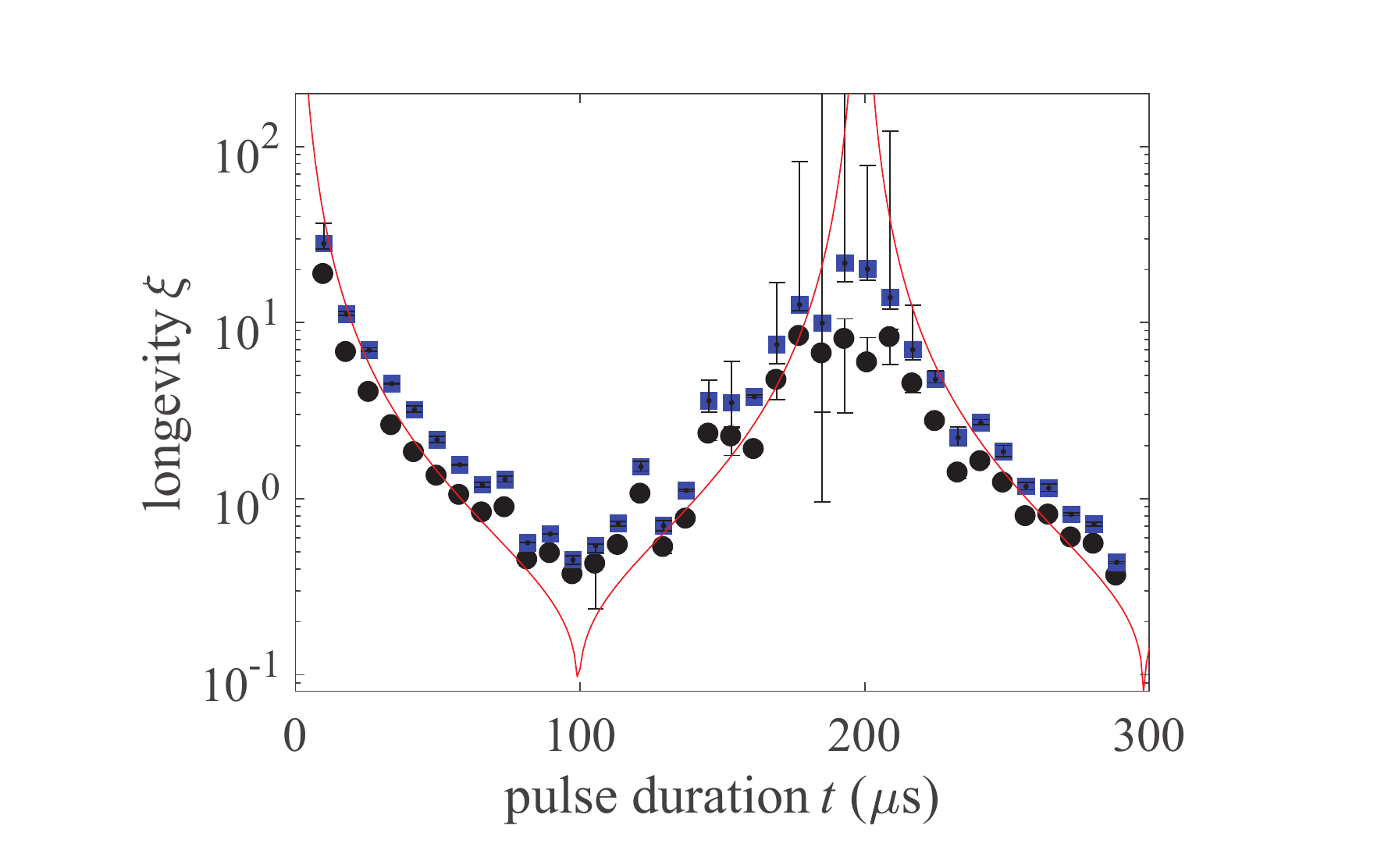} &
		\includegraphics[width=0.455\linewidth]{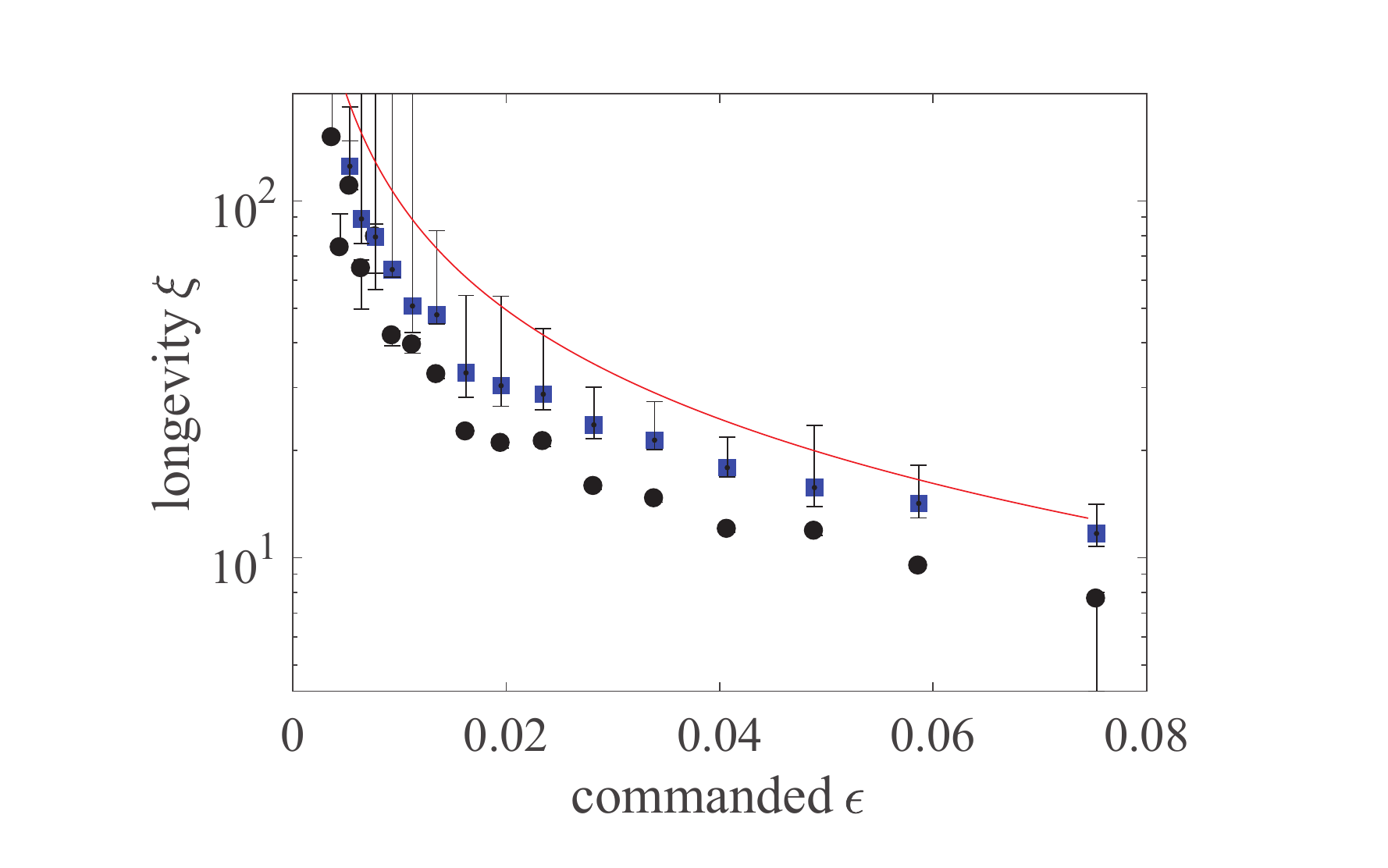}	
		\end{array}$
		\caption{The tuneability of $\xi$. The red curves depict the expected values of $\xi$ based on the model in Section \ref{tech}. The black circles depict the value of $\xi$ obtained from an exponential fit as shown in Fig. \ref{figEXP}b, and the blue squares depict the values of $\xi$ obtained from a fit to a model which accounts for atomic recoil after coherent transfer, as described in Eq. \ref{rescatt}, with $\beta = 0.0091$. Error bars represent 95 \% confidence intervals. \textbf{(a)} The longevity $\xi$ varies as a function of the duration of the microwave pulse. \textbf{(b)} The longevity $\xi$ varies as a function of the fraction transferred. Here the longevity is tuned by varying the amplitude of the microwave pulse.}
		\label{fig3}
	\end{center}
\end{figure}

Another comparison of the two models can be seen in Fig. \ref{fig3}, where the red curves depict the expected values of $\xi$ based on the commanded $\epsilon$ as described in Section \ref{tech}, the black circles depict the measured $\xi$ determined from a simple exponential fit as in Eq. \ref{expparam}, and the blue squares depict the measured $\xi$ once the data is fit to Eq. \ref{rescatt} rather than a simple exponential. For the exponential model, we fit the atom number in a series of PTAI shots to an exponential function and used the decay constant to obtain $\xi$ as in Eq.\ref{expparam}. For the rescattering model, we fit the atom number in a series of PTAI shots to Eq. \ref{rescatt} with $\epsilon$ as a free parameter and obtained $\xi=-1/\mathrm{ln} \left(1-\epsilon\right)$ as in Eq. \ref{expparam}. Analyzing the data using the rescattering model more often produces the expected value of the longevity, especially for high values of $\xi$, or, equivalently, low values of $\epsilon$, as shown in Fig. \ref{fig3}b.

The two panels in Fig. \ref{fig3}b depict different ways of tuning the longevity. In Fig. \ref{fig3}a, we tuned $\xi$ by varying the microwave pulse duration as described in Section \ref{tech}. In Fig. \ref{fig3}b, we tuned $\xi$ by varying the microwave pulse amplitude, such that higher degrees of pulse attenuation decreased the commanded transfer fraction $\epsilon$. The fact that the longevity can be tuned in different ways illustrates the versatility of the partial-transfer imaging technique for implementation in various experiments.

%% FIGURE 5

\section{Application: Collective Modes} \label{dipolemode}

Partial-transfer absorption imaging is very useful for studies in which it is important to sample the same atomic clod repeatedly. The observation of collective modes in a BEC --- for example, to measure trap frequencies --- is one example of an application for PTAI. We measured our dipole trap frequency by exciting and measuring the dipole mode of the BEC along ${\bf e}_{x}$.

%% FIGURE 6
\begin{figure*}[ht]
	\begin{center}
		\includegraphics[scale=0.32]{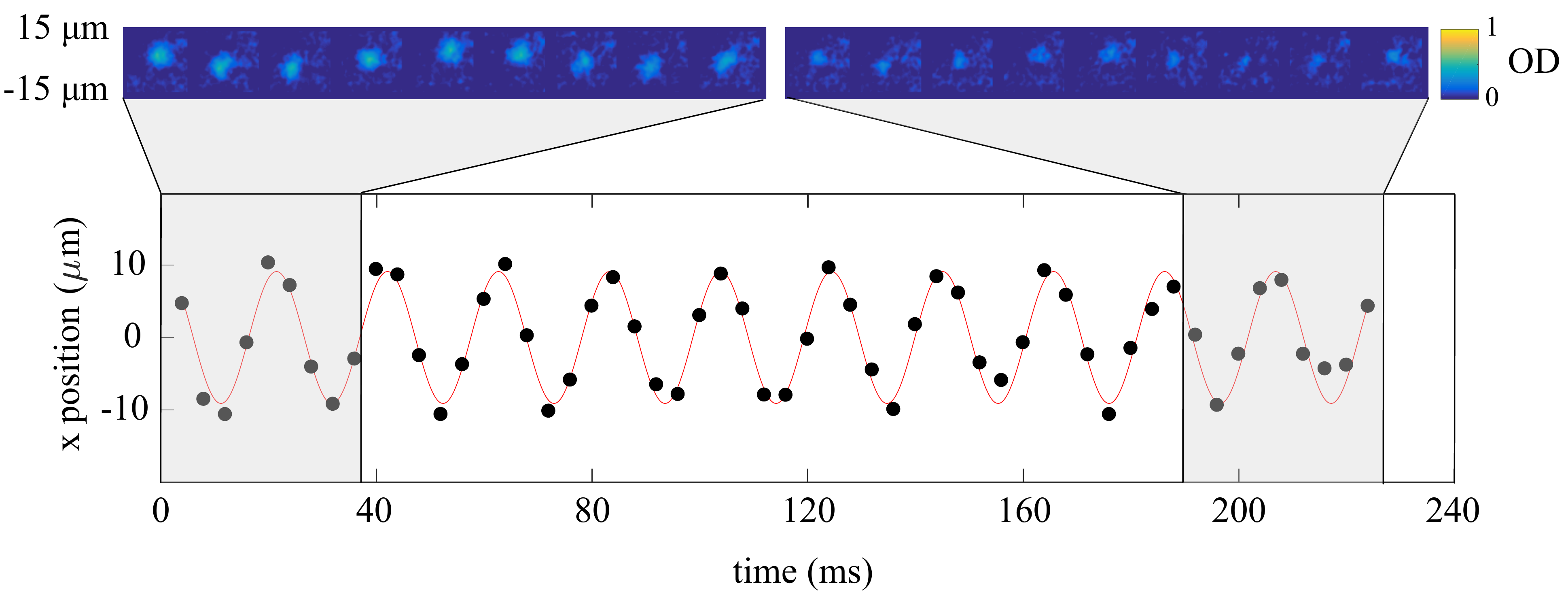}
		\caption
		{  A collective dipole excitation in a BEC. A series of spatially low-passed images of the same BEC are inset at the top of the figure, showing a dipole mode. For ease of viewing, only the first and last 9 PTAI shots are shown. The center of mass position of the BEC along $\mathbf{e}_{x}$ is plotted as a function of PTAI shot number. The data show a clear oscillation without damping.
		}
		\label{fig6}
	\end{center}
\end{figure*}

We displaced our optical dipole trap center along ${\bf e}_{x}$ by laterally displacing one of the dipole beams, then abruptly restored it to its original position and allowed the BEC to oscillate in the trap. We repeatedly applied microwave pulses of duration 7 $\micro$s, transferring 7.5 \% of the atoms from $\ket{1,-1}$ to $\ket{2,-2}$, and imaged the BEC 56 times in order to observe the dipole mode. The PTAI images were separated in time by 4 ms, as shown in Fig. \ref{fig6}. Fitting to the center of mass of each of the images gives an approximately 50 Hz trap frequency along ${\bf e}_{x}$ (parallel to one of our dipole trap laser beams).

Before finding the center of mass of the BEC in Fig. \ref{fig6}, we applied a lowpass spatial filter to the images by convolving them with a Gaussian kernel \cite{Crocker1996}. The lowpass filter mitigates the effect of shot noise while preserving integrated atom number. We chose the cutoff spatial period to be smaller than the resolution of the imaging system to avoid discarding information about the cloud. We then found the center of mass by fitting the lowpassed images to a Thomas-Fermi profile and taking the peak position to be the center of mass.

As can be seen from Fig. \ref{fig6}, the resulting center of mass positions show nearly undamped oscillations, while the atom number is significantly depleted by the end of the measurement sequence. Because of the ready ability to locate the BEC's position even at the end of the measurement sequence, it is clear that the center of mass measurement could have been performed with far weaker measurements, and therefore increased longevity.

\section{Discussion}

In every minimally-destructive imaging technique, there is a trade-off between image signal strength and the number of images of the cloud that can be captured. In PTAI the trade-off is adjusted by controlling the parameters of a pulse which couples bright and dark states rather than the optical parameters affecting the atom-light interaction. The selected duration and amplitude of the coupling pulse determine the fraction of atoms that is transferred into the bright state and subsequently imaged in each of a series of images. Larger transfer fractions lead to fewer images, each with larger signal, while smaller transfer fractions lead to a larger number of images, each with with reduced signal.

Multi-shot partial-transfer absorption imaging, though valuable, is not without its limitations. Measurements of atomic clouds of high OD typically underrepresent the optical depth of the cloud, which is, of course, why time-of-flight is such a common technique for imaging optically thick clouds. Additionally, rescattering of atoms in $\ket{d}$ by atoms in $\ket{b}$ reduces the reproducibility of the measurements. The PTAI technique is best-suited for small $\epsilon$, and for those who seek out a minimally-destructive imaging technique, many weak measurements are likely to be preferable over a few measurements with enhanced signal.

We have shown that multi-shot PTAI is a valuable technique for repeated measurements of an atomic cloud and that the strength of the measurements can be easily tuned. In the case of strong transfer (high $\epsilon$ and low $\xi$), a few relatively high-signal images of the BEC may be captured before the cloud is depleted. For weak measurements (low $\epsilon$ and high $\xi$) the same ensemble may be sampled tens of times with minimal perturbation to the cloud.

\section{Acknowledgements}

This work was partially supported by the AROs atomtronics MURI, the AFOSRs Quantum Matter MURI, NIST, and the NSF through the PFC at the JQI.

\appendix

\section{Calculation of I$_\mathrm{sat}$ and $\alpha$}

The values for the parameters $I_\mathrm{sat}$ and $\alpha$ from Equation \ref{nsig0} were calculated for the Princeton Instruments ProEM HS: 512BX3 camera we used. The camera has a 512$\times$512 imaging array with $16 \micro{\rm m} \times 16 \micro {\rm m}$ square pixels. We hardware-binned the pixels into $3 \times 3$ groups, giving effective $48 \micro {\rm m} \times 48 \micro {\rm m}$ super-pixels.

%% FIGURE 7
\begin{figure}[tbph]
	\begin{center}
		$\begin{array}{ll}
		\textrm{\textbf{(a)}} & \textrm{\textbf{(b)}} \\ \includegraphics[width=0.4\linewidth]{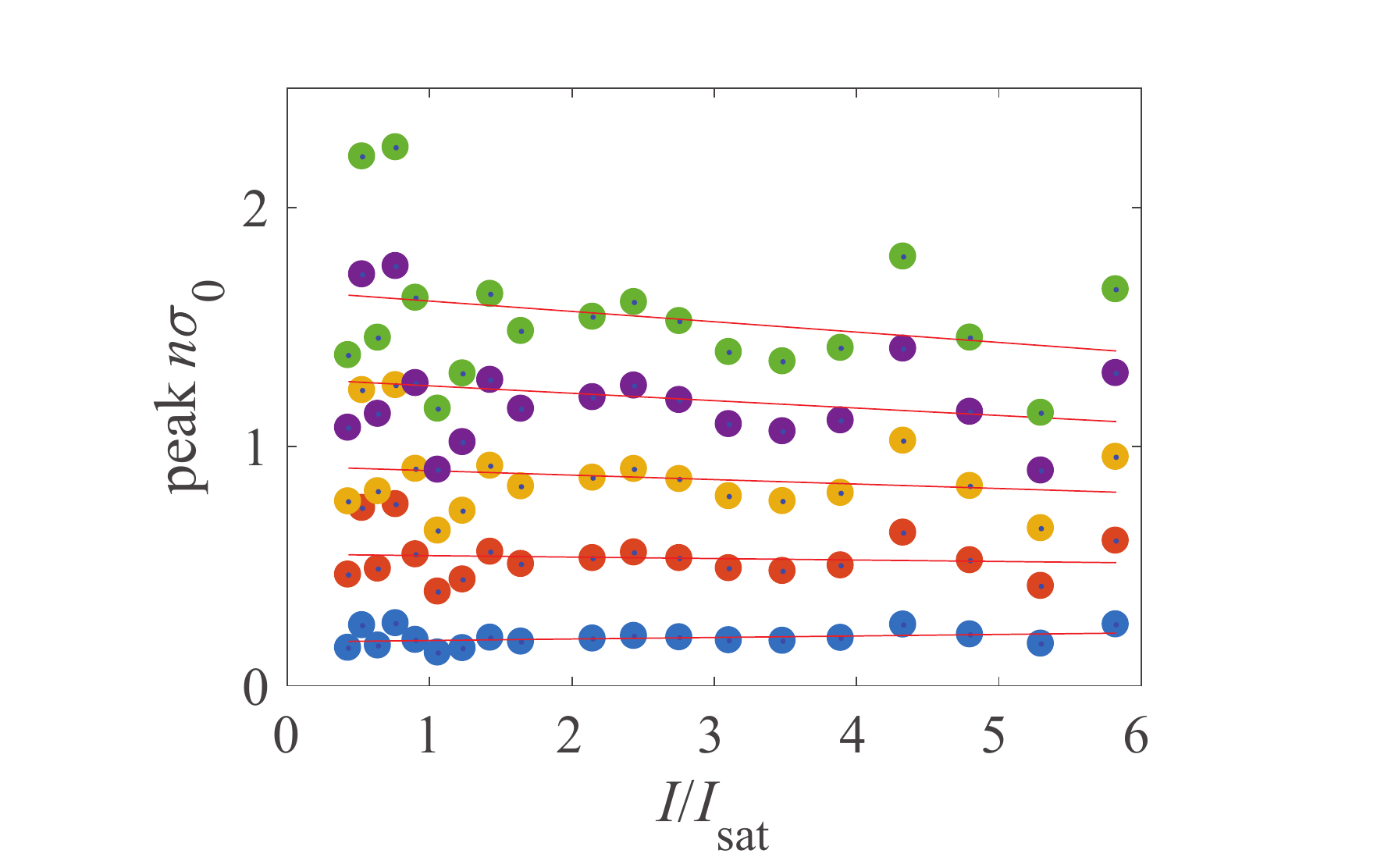} & 
			\includegraphics[width=0.45\linewidth]{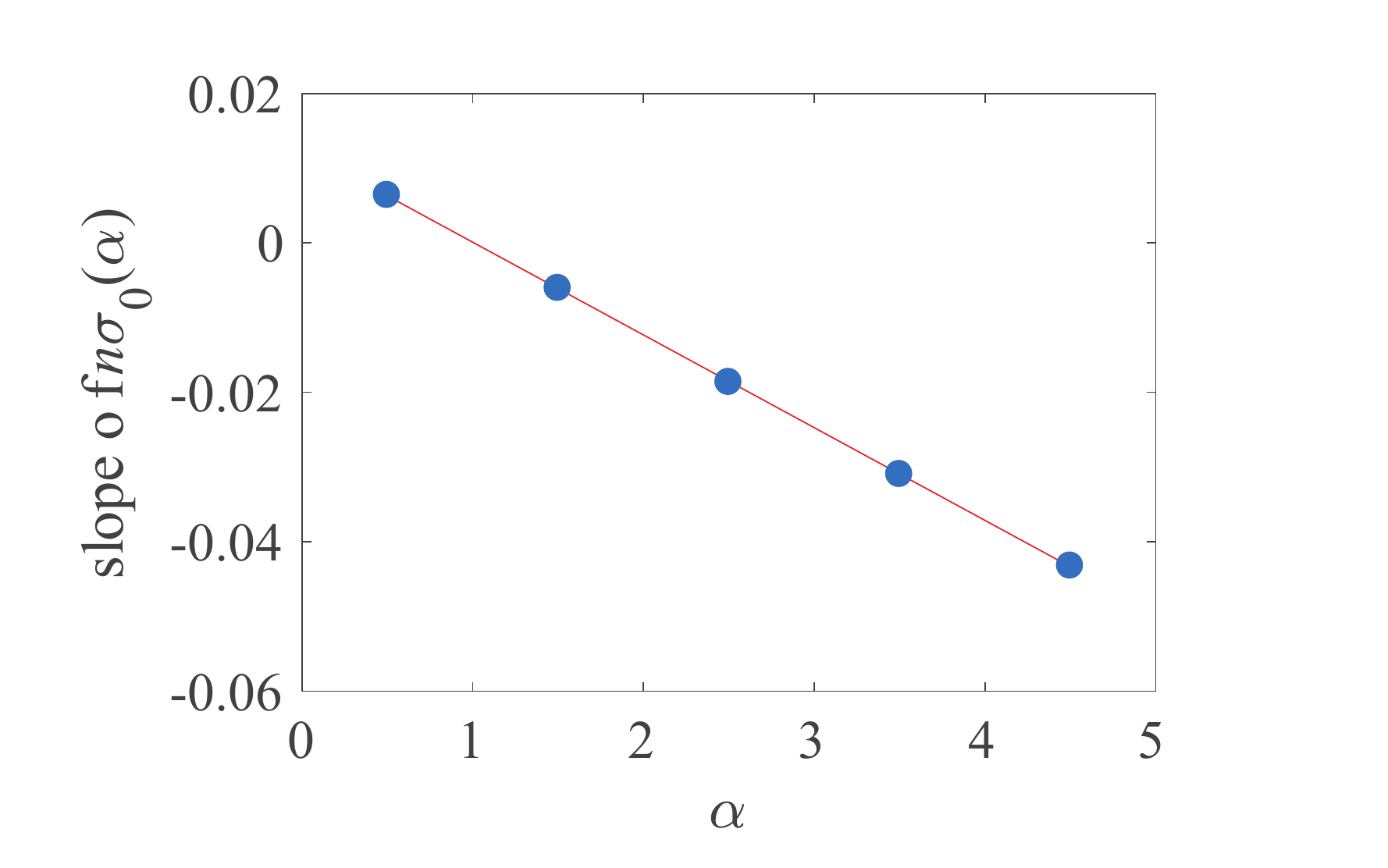}
		\end{array}$
		\caption
		{  \textbf{(a)} The series of curves each shows $n \sigma_{0}$ as a function of $I/I_{\mathrm{sat}}$ for a candidate value of $\alpha$ along with a linear fit. \textbf{(b)} Here the slope of the linear fit to each $n \sigma_{0} (\alpha)$ curve is shown as a function of the candidate value of $\alpha$. The value chosen for $\alpha$ is where the slope goes to zero.
		}
		\label{fig7}
	\end{center}
\end{figure}

At our 780 nm imaging transition, the 1.669(2) mW/cm$^{2}$ saturation intensity corresponds to a 6.551(8) $\times 10^{16}$ photon/s/cm$^{2}$ flux. For our 25 $\micro {\rm s}$ imaging pulse duration, a conversion of 0.38 photo-electrons per 16-bit analog-digital unit (ADU), and the camera's 78 \% quantum efficiency, each CCD super-pixel registers about 4300 ADUs when illuminated at $I_{\rm sat}$. Given an optical transmission of 80 \%, this number is reduced to about 3400 counts per pixel.

Following the procedure in Ref. \cite{Reinaudi2007}, we obtained $\alpha$ by first computing $n \sigma_{0}$ as a function of $I/I_{\mathrm{sat}}$ for a variety of candidate values of $\alpha$ from 0.5 to 5, as shown in Fig. \ref{fig7}a. Since the physical product $n \sigma_{0}$ is independent of $I$, we plotted the computed $n \sigma_{0}$ as a function of $I/I_{\mathrm{sat}}$ and fitted each curve to a line. Plotting the slopes of the $n \sigma_{0}$ curves as a function of $\alpha$, as shown in Fig. \ref{fig7}b, we selected the value of $\alpha$ which gave a slope of zero. This analysis yielded $\alpha$ = 1.13(2).

\bibliography{PTAI}

\end{document}